\documentclass[sn-mathphys,Numbered]{sn-jnl}

\usepackage{graphicx}%
\usepackage{multirow}%
\usepackage{amsmath,amssymb,amsfonts}%
\usepackage{amsthm}%
\usepackage{mathrsfs}%
\usepackage[title]{appendix}%
\usepackage{xcolor}%
\usepackage{textcomp}%
\usepackage{manyfoot}%
\usepackage{booktabs}%
\usepackage{algorithm}%
\usepackage{algorithmicx}%
\usepackage{algpseudocode}%
\usepackage{listings}%
\usepackage{braket}

\theoremstyle{thmstyleone}%
\theoremstyle{thmstyletwo}%

\theoremstyle{thmstylethree}%

\begin{document}

\title[Article Title]{Machine Learning Methods for Background Potential Estimation in 2DEGs}

\author*[1]{\fnm{Carlo} \sur{da Cunha}}\email{carlo.cunha@nau.edu}
\author[2]{\fnm{Nobuyuki} \sur{Aoki}}
\author[3]{\fnm{David} \sur{Ferry}}
\author[4]{\fnm{Kevin} \sur{Vora}} 
\author[4]{\fnm{Yu} \sur{Zhang}}
\affil*[1]{\orgdiv{School of Informatics, Computing, and Cyber-Systems}, \orgname{Northern Arizona University}, \orgaddress{ \city{Flagstaff}, \postcode{86011}, \state{AZ}, \country{USA}}}
\affil[2]{\orgdiv{Department of Electronics and Mechanical Engineering}, \orgname{Chiba University}, \orgaddress{\city{Chiba}, \postcode{263-8522}, \state{Chiba}, \country{Japan}}}
\affil[3]{\orgdiv{School of Electrical, Computer and Energy Engineering and Center for Solid State Electronics Research}, \orgname{Arizona State University}, \orgaddress{\city{Tempe}, \postcode{85287}, \state{AZ}, \country{USA}}}
\affil[4]{\orgdiv{School of Computing and Augmented Intelligence}, \orgname{Arizona State University}, \orgaddress{\city{Tempe}, \postcode{85281}, \state{AZ}, \country{USA}}}


\abstract{In the realm of quantum-effect devices and materials, two-dimensional electron gases (2DEGs) stand as fundamental structures that promise transformative technologies. However, the presence of impurities and defects in 2DEGs poses substantial challenges, impacting carrier mobility, conductivity, and quantum coherence time. To address this, we harness the power of scanning gate microscopy (SGM) and employ three distinct machine learning techniques to estimate the background potential of 2DEGs from SGM data: image-to-image translation using generative adversarial neural networks, cellular neural network, and evolutionary search. Our findings, despite data constraints, highlight the effectiveness of an evolutionary search algorithm in this context, offering a novel approach for defect analysis. This work not only advances our understanding of 2DEGs but also underscores the potential of machine learning in probing quantum materials, with implications for quantum computing and nanoelectronics. }




\maketitle

\section{Introduction}\label{sec1}

Quantum-effect devices and materials, characterized by exotic quantum properties, may hold the foundation for next-generation technologies. Within this landscape, two-dimensional electron gases (2DEGs) emerge as fundamental structures, often crafted through the use of two-dimensional materials \cite{CRC2016_1,CRC2017_1} or heterostructures \cite{CRC2005_1}. They play an important role in novel technologies such as quantum computing, providing a precisely controlled environment to exploit quantum phenomena. Moreover, 2DEGs show intriguing phenomena such as the quantum Hall effect \cite{QHall}, with direct applications in precision metrology \cite{Metrology}.

The presence of impurities and defects can significantly impact the electronic properties of 2DEGs \cite{dis1,dis2}. These anomalies introduce energy levels within the bandgap and scatter electrons, directly influencing carrier mobility and conductivity. High-electron-mobility transistors (HEMTs) \cite{HEMT} and other devices rely on precise control of carrier concentration and mobility, both of which are susceptible to defects and impurities \cite{CRC2006_4}. Uncontrolled defects can also introduce decoherence, limiting the quantum coherence time of qubits and raising reliability concerns in electronic devices.

However, detecting the background potential caused by these defects and impurities poses a considerable challenge. 2DEGs are remarkably thin, and defects often occur at the nanometer scale or smaller. Moreover, these imperfections may not manifest as visible features but can exert only indirect effects on electronic properties. 

Scanning gate microscopy (SGM) \cite{Westervelt} emerges as a valuable technique to address this challenge \cite{CRC2014_2}. SGM employs a sharp charged tip to perturb a 2DEG while concurrently monitoring its conductivity. This yields a conductance map across the device, reflecting carrier flow in response to the perturbation. This technique has been used to image compressibility features of the quantum Hall regime \cite{CRC2005_2}, to estimate ballistic carrier trajectories in graphene \cite{SGMBal}, and to visualize wavefunction scars in quantum dots \cite{WVScar}. While the information provided by SGM may require nuanced interpretation \cite{Hard1,Hard2,Hard3}, it can be leveraged to indirectly infer details about the background potential \cite{SGMP1,SGMP2,SGMP3}.

In this work, we undertake a comparative study of three machine learning techniques applied to estimate the background potential of a 2DEG using SGM data.
To focus on a well-defined and localized region of the 2DEG, we utilize a quantum point contact (QPC) \cite{Wees} as a prototypical device \cite{CRC2007_2,CRC2006_1,CRC2006_2}. QPCs find wide-ranging applications in nanoelectronics and quantum computing, underscoring the significance of understanding the background potential for optimizing device performance.

Our estimation of the background potential from the SGM conductance map hinges on three different techniques: a picture-to-picture translator based on generative adversarial neural networks (Pix2Pix GAN), cellular neural networks (CNN) \cite{CRC2022_1,CNN,CNN2}, and an evolutionary search algorithm (ES) \cite{CRC2023_1}. Other attempts to solve this inverse problem include the use of deep convolutional neural networks \cite{ladrao1} and deep reinforcement learning \cite{ladrao2}. In all techniques, a conductance map is obtained from SGM, and the inverse problem of estimating the background potential is solved using different machine learning techniques.

We gauge the performance of these techniques through a series of calculated parameters. First, we calculate the entropy of the estimated potential images. This is a technique commonly used to assess the quality of microscopy images. For example, images with high entropy values contain more information, while images with low entropy values are often associated with uniform and repetitive patterns \cite{Entrop1,Entrop2}. The spatial variability of the estimated potential is studied with surface roughness analysis, while the fractal dimension offers information about the potential at different scales. Fractal dimension studies appear as an important concept in quantum-effect devices. For instance, it is used to study signatures of chaotic behavior in quantum dots \cite{QChaos1,QChaos2,CRC2016_1}. Complementing these techniques, we also studied the distribution of sizes of potential features using morphological granulometry. The importance of performing a granulometric analysis relies on the fact that variations in the potential can impact the behavior of the device significantly. Finally, we also estimate the structure of a quenched structure of the potential to identify static features such as those associated with lattice defects.

Our results, drawn from a limited pool of experimental data, suggest that, given the constraints of data availability and acquisition costs, the evolutionary search technique yields the most promising outcomes.

This study showcases the effectiveness of machine learning techniques in estimating the background potential of 2DEGs, offering valuable insights that can guide future research in addressing similar inverse problems.

The structure of this manuscript is outlined as follows: First we describe the experimental methods to fabricate the 2DEG, the QPC and how the SGM measurement were obtained. We then describe how we used Green's functions \cite{Green1} and perturbation theory \cite{CRC2022_1} to obtain image pairs of different potential configurations and the corresponding SGM signal. Next we describe the three machine learning techniques: Pix2Pix GAN, CNN, and ES. We then show the results for image entropy, roughness, fractal dimension, and morphological granulometry. We complete the discussion of the results with an estimate of the quenched disorder potential.

\section{Methods}\label{sec:methods}

\subsection{Experimental}\label{sec:Exp}
A two-dimensional electron gas (2DEG) was created by a modulation-doped heterostructured consisting of In$_{0.53}$Al$_{0.47}$As/In$_{0.53}$Ga$_{0.47}$As/In$_{0.53}$Al$_{0.47}$As layers on an (001)  InP substrate \cite{CRC2005_1,CRC2006_4,CRC2016_1}. This 2DEG is located $45$ nm below the surface, which is coated with $30$ nm of polymethylmethacrylate (PMMA) to avoid short-circuits between the tip and the device. 

\begin{figure}
	\centering
	\includegraphics{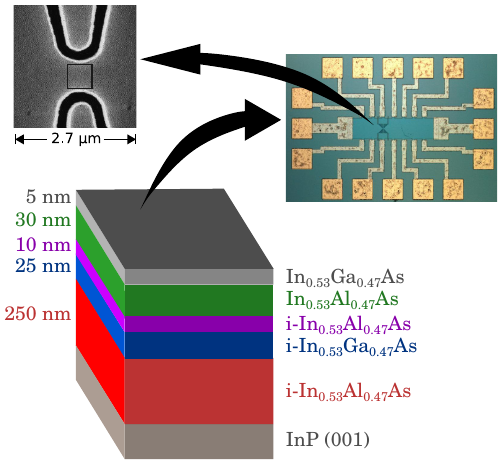}
	\caption{A sketch of the InGaAs/InAlAs heterostructure, a Hall bar defined lithographically on its surface, and the quantum point contact etched on the Hall bar.}
	\label{fig:QPC}
\end{figure}

Shubnikov-de Haas measurements revealed the occupation of two subbands with electron densities of $7.2\times 10^{11}$ cm$^{-2}$ and $2.1\times 10^{11}$ cm$^{-2}$. This corresponds to an electron mobility of $7.4\times 10^4$ cm$^2$/V$\cdot$s and a mean-free path of $1.2$ $\mu$m \cite{CRC2005_1,CRC2006_1,CRC2006_4,CRC2007_2}. A quantum point contact (QPC) was etched on top of the heterostructure using electron beam lithography. The QPC has a physical opening of $600$ nm and the etched trenches have a radius of approximately $800$ nm. A voltage between $0$ V and $-7.2$ V was applied to the in-plane gates defined by the etched trenches to control the electrical width of the QPC. The material structure as well as a Hall bar lithographically defined on its surface, and the QPC are shown in Fig. \ref{fig:QPC}.

Scanning gate measurements were conducted by rastering a tip coated with $15$ nm of PtIr $40$ nm above the PMMA layer while changes in conductance were monitored. A current of $10$ nA was applied across the QPC, while the voltage was measure with a lock-in amplifier operating in quasi-DC. All measurements were performed at a system temperature of $280$ mK, and the tip was biased so that the maximum change in conductance was $0.2\times 2e^2/h$.

\subsection{Computational}\label{sec:Comp}
Three distinct methods for estimating the background potential from SGM data were used: i) Pix2Pix GAN, ii) Cellular Neural Networks, and iii) Evolutionary Search.
\subsubsection{Green's Functions}\label{sec:Green}
In the weak-perturbation regime, the tip potential shifts the eigenstates of the system according to first-order perturbation theory \cite{CRC2022_1}:

\begin{equation}
	\Delta\varepsilon_n=\braket{n|V|n},
\end{equation}
where $n$ is an eigenstate, and $V$ is the tip-induced potential. If the tip-induced potential is sufficiently sharp, it can be approximated by:

\begin{equation}
	V(\mathbf{r})=V_0\delta(\mathbf{r}-\mathbf{r}_0),
\end{equation}
where $\mathbf{r}_0$ is the position of the tip on the plane of the 2DEG. This leads to a shift in eigenenergy given by:

\begin{equation}
	\Delta\varepsilon_n=V_0|\varphi_n(\mathbf{r}_0)|^2,
\end{equation}
where $\varphi_n(\mathbf{r}_0)$ is the wavefunction evaluated at $\mathbf{r}_0$.

Consequently, changes in conductance are given by:

\begin{equation}
	\Delta G(\mathbf{r}_0;\varepsilon)\approx\frac{\partial G}{\partial\varepsilon_F}\sum_n\Delta\varepsilon_n\delta(\varepsilon-\varepsilon_n)\propto D(\mathbf{r}_0).
\end{equation}
Here, we specifically focus on states indexed by $n$ that are  in proximity to the Fermi energy, and introduce the concept of the local density of states (LDOS), denoted by $D$.

The tip-induced potential on the 2DEG, however, is given by:

\begin{equation}
	V_{tip}(\mathbf{r},\mathbf{r}_{tip})=\frac{d_0^2}{\|\mathbf{r}-\mathbf{r}_{tip}\|^2+d_0^2}.
\end{equation}
In this expression, the parameter $d_0$ is held contant at a value corresponding to 80 lattice constants. Additionaly, $\mathbf{r}_{tip}$ signifies the location of the tip within the plane of the 2DEG. We apply a truncation to the tip potential, limiting its influence to a radius that aligns closely with $14$ lattice constants, a value approximating the physical tip radius.

Consequently, changes in conductance, in the weak-perturbation regime, can be approximated by:

\begin{equation}
	\Delta G(\mathbf{r}_0;\varepsilon)\approx\sum_{\mathbf{r}}V_{tip}(\mathbf{r},\mathbf{r}_{tip})D(\mathbf{r}).
\end{equation}	
To find $D(\mathbf{r})$, a potential is generated, the corresponding Hamiltonian $\mathbf{H}$ is calculated, and then the corresponding Green's function \cite{Green1} is found. Instead of computing the Green's function for the entire potential, the images were segmented into slices, and the Green's function was computed to each slice $n$ using:

\begin{equation}
	\mathbf{G}_{nn}(\varepsilon)=\left[\left(\varepsilon+i\eta\right)\mathbf{I}-\mathbf{H}_n\right]^{-1},
\end{equation}
where $\eta\rightarrow 0$ ensures that the retarded Green's function is obtained. The slices are then merged using Dyson's equation:

\begin{equation}
	\mathbf{G}_{ab}=\mathbf{G}^0_{ab}+\sum_{mn}\mathbf{G}^0_{am}\mathbf{V}_{mn}\mathbf{G}_{nb},
\end{equation}
where the upscript $0$ indicates the unperturbed Green's function for the slice, and $\mathbf{V}$ is the coupling matrix between slices.
The local density of states (LDOS) is given by:

\begin{equation}
	D_{mn}=-\frac{1}{\pi}Im\left[G_{nn;mm}\right].
\end{equation}

Given the computed carrier density, the Fermi wavelength is $\sqrt{2\pi/n_{2D}}\approx 30$ nm, which is comparable to the physical tip radius.
All images were computed on a $600\times 600$ grid, which corresponds to an area of $675\times 675$ nm$^2$. This implies a discretization spacing of $1.125$ nm, which gives a wavenumber of $0.21$ nm$^{-1}$, which is one order of magnitude away from the first Brillouin zone located at $5.58$ nm$^{-1}$. A saddle potential was included to mimic the behavior of the side-gate potential.

\subsubsection{Pix2Pix GAN}\label{sec:GAN}
Spatial variations in the background potential should be close to the Bohr radius ($4\pi\varepsilon\hbar^2/m^*e^2$) \cite{InGaAs}, which, for InGaAs, is approximately $18$ nm, corresponding to eight lattice constants.
Random potentials were created, where each site is occupied with a probability $p\in[0,0.04]\%$. If the site is occupied, then a truncated Coulomb-like potential is created around a radius of five lattive constants. The maximum potential at the site is $10$ meV. This impurity representation fully captures the expected variations in the background potential. A total of $1600$ potentials were created, and, using image augmentation corresponding to flipping and salt-and-pepper noise, this number was increased to $12800$.

The Pix2Pix GAN \cite{Pix2Pix} is a deep learning architecture that learns to translate images from one domain to another, such as converting sketches into realistic images. Given our dataset of input images, we would like Pix2Pix GAN to predict the target images available in the output dataset. Next, we describe the Pix2Pix GAN and its underlying concepts, including the generator and discriminator networks. The Pix2Pix model is a type of conditional GAN, or cGAN, where the generation of the output image is conditional on input. The discriminator is provided both with an input image and the target image and must determine whether the target is a plausible transformation of the input image while the generator is responsible to generate the target images.

The generator is an encoder-decoder model which uses a U-Net architecture \cite{UNet}. The model takes an input image and generates a target image. It does this by first downsampling or encoding the input image down to a bottleneck layer, then upsampling or decoding the bottleneck representation to the size of the output image. The U-Net architecture means that skip-connections are added between the encoding layers and the corresponding decoding layers, forming a U-shape. The generator is trained via adversarial loss, which encourages the generator to generate plausible images in the target domain. The generator is also updated via the mean absolute error (MAE) loss measured between the generated image and the expected output image. This additional loss encourages the generator model to create plausible translations of the input image. 

The discriminator is a deep convolutional neural network that performs image classification. Specifically, conditional-image classification. It takes both the input image and the target image as input and predicts the likelihood of whether the target image is real or a fake translation of the source image. The discriminator design is based on the effective receptive field of the model, which defines the relationship between one output of the model to the number of pixels in the input image. This is called a PatchGAN model and is carefully designed so that each output prediction of the model maps to a $70\times 70$ square or patch of the input image. The benefit of this approach is that the same model can be applied to input images of different sizes, e.g. larger or smaller than $256\times 256$ pixels. The output of the model depends on the size of the input image but may be one value or a square activation map of values. Each value is a probability for the likelihood that a patch in the input image is real. These values can be averaged to give an overall likelihood or classification score if needed. The discriminator is updated in a standalone manner, so the weights are reused in this composite model but are marked as not trainable. The composite model is updated with two targets, one indicating that the generated images were real (cross entropy loss), forcing large weight updates in the generator toward generating more realistic images, and the executed real translation of the image, which is compared against the output of the generator model (L1 loss).

Implementation Details: We first pre-process the data and implement a Keras (with TensorFlow backend) version of the original implementation\footnote{Available at https://github.com/phillipi/pix2pix}. Each image is loaded, rescaled, and split into the input and target elements. Each image pairs are resized to the width and height of $256\times 256$ pixels. Once loaded, we can save the prepared arrays to a new file in compressed format for later use. Typically, GAN models do not converge; instead, an equilibrium is found between the generator and discriminator models. As such, we cannot easily judge when training should stop. Therefore, we can save the model and use it to generate sample image-to-image translations periodically during training, such as every $10$ training epochs. We can then review the generated images at the end of training and use the image quality to choose a final model. We run for approximately $100$ epochs for the presented results.

\subsubsection{Cellular Neural Networks}\label{sec:CNN}
Images are segmented into $N\times N$ discrete cells whose states $\{X_{\mathbf{i}}\mid i_{x,y}=1,2,\hdots,N\}$ evolve according to \cite{CNN}:

\begin{equation}
	\frac{dX_\mathbf{i}(t)}{dt}=-X_\mathbf{i}(t)+\sum_{\mathbf{j}\in\mathcal{N}(\mathbf{i})}A_{\mathbf{ij}}Y_\mathbf{j}(t)+\sum_{\mathbf{j}\in\mathcal{N}(\mathbf{i})}B_{\mathbf{ij}}U_\mathbf{j}(t) + Z,
	\label{eq:CNN}
\end{equation}
where $\mathbf{A}$ and $\mathbf{B}$, known as cloning templates, are tensors used to propagate specific information behaviors within a local neighborhood of cells across the image.
 $\mathbf{U}$ is an input to the cell, $Z$ is an offset, typically set to $1$, and $\mathbf{Y}$ is the output of the cell, given by $Y_{\mathbf{i}}(t)=\text{tanh}\left(X_{\mathbf{i}}(t)\right)$. Also, $\mathcal{N}(\mathbf{i})$ denotes the neighborhood of cell $\mathbf{i}$. The Moore neighborhood, defined as $\mathcal{N}_{\mathbf{i}}=\{\mathbf{j}: |i_x-j_x|\leq 1,\ |i_y-j_y|\leq 1\}$, was used in all computations.

To obtain the steady-state output of the CNN, we discretize Eq. \ref{eq:CNN} and use Euler's method to obtain update rule for the cell states:

\begin{equation}
	\mathbf{X}^{n+1}=(1-\Delta)\mathbf{X}^n+\Delta\boldsymbol{\xi}^n,
\end{equation}
where $\Delta=10^{-3}$ is a unit time period, and:

\begin{equation}
	\boldsymbol{\xi}^n=\mathbf{A}\cdot\mathbf{Y}^n+\mathbf{B}\cdot\mathbf{U}+\mathbf{J}.
\end{equation}
In this last equation, the dot represents the contraction between a tensor and a multi-dimensional array. $\mathbf{J}$ is a full-rank array of ones. The dynamical system was evolved until the maximum difference between $\mathbf{X}^{n+1}$ and $\mathbf{X}^n$ was less than $10^{-3}$.

For training, the input $\mathbf{U}$ is set to the theoretically computed SGM image, while the expected steady-state output is the potential used to compute the corresponding SGM image. The dynamical system is evolved until steady-state, and then the cloning templates are updated using gradient ascent according to:

\begin{equation}
	\mathbf{A}^{n+1},\mathbf{B}^{n+1}=\mathbf{A}^n,\mathbf{B}^n-\eta\nabla_{\mathbf{A},\mathbf{B}} L(\mathbf{A}^n,\mathbf{B}^n),
\end{equation}
where $\eta=10^{-3}$ is the learning rate, and $L$ is the loss, defined as the correlation between the obtained state $\mathbf{X}$ and the expected experimental SGM image.

The number of images and their preparations were the same used in Pix2Pix GAN.

\subsubsection{Evolutionary Search}\label{sec:swarming}
The evolutionary algorithm \cite{CRC_ML_Book,CRC2023_1} starts with a population of $\{\mathbf{U}_i\mid i=1,2,\hdots,20\}$ individual corresponding to different potentials. At each interation, the SGM images corresponding to each test solution are computed together with the correlation $r_i$ between them and the experimental SGM image. The individual with the highest correlation is declared the winner, and the rest of the swarm move towards it according to:

\begin{equation}
	\mathbf{U}_i^{n+1}=\alpha\mathbf{U}_i^n+(1-\alpha)\mathbf{U}_{winner}^n+\theta(r_i)\boldsymbol{\eta},
\end{equation}
where $\alpha=0.95$ is a hyperparameter, $\boldsymbol{\eta}\sim\mathcal{N}(0,1)$, and $\theta(r)$ is given by:

\begin{equation}
	\theta(r)=\left\{\begin{array}{cl}
		\theta_0e^{-\gamma r}&\text{if } r>0,\\
		\theta_0&\text{otherwise},
	\end{array}\right.
\end{equation}
where $\theta_0=0.1$ meV and $\gamma=5$. This function garantees that the adjustment is small if the correlation is high, mimicking the dynamics of simmulated anneling. 

This relaxation procedure was performed until the highest correlation was at least $72\%$. Although this may be considered a threshold, it takes a considerably long computation time using current technology. Nonetheless, this correlation between SGM signals corresponds to a correlation of $78\%\pm10\%$ between the estimated and expected potentials for Gaussian fields.

\section{Results}\label{sec:results}

\begin{figure}[h!]
	\centering
	\includegraphics[scale=0.75]{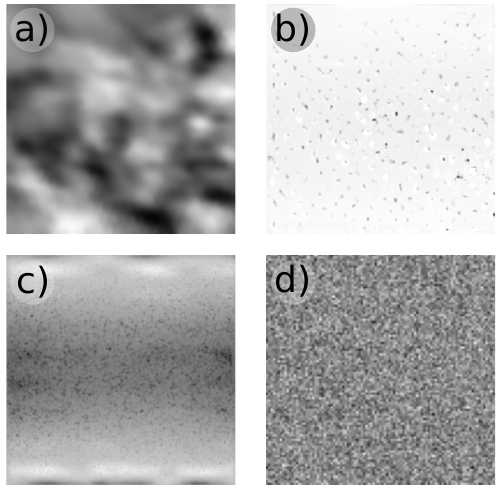}
	\caption{a) Experimental SGM image and the estimated background potential using b) Pix2Pix GAN, c) CNN, and d) Evolutionary Search. All images correspond to an area of $675\times 675$ nm$^2$.}
	\label{fig:comparative}
\end{figure}

Figure \ref{fig:comparative} shows a typical SGM image and the corresponding background potential using the techniques studied in this work. Surprisingly, the background estimates exhibit notable discrepancies depending on the employed method. 
Since no ground truth is available, finding the method that gives the most appropriate result becomes a formidable task.

\subsection{Image Entropy}
To quantify the estimation difference, we computed the image entropy for the whole set of images for each estimation technique. The image entropy is defined as:

\begin{equation}
	S=-\sum_{\langle image\rangle}p_i\log_2(p_i),
\end{equation}
where $p_i$ is the probability of finding a pixel of the image with intensity $i$. This probability was computed from the histogram of the image using $256$ bins. 

The importance of using the image entropy to qualify differences in estimation relies on the fact that the image entropy quantifies the level of complexity in the distribution of the intensities within the image. For example, if it is equally probable to find any intensity in a binary image, $p_i=1/2$ and the entropy is one, representing complete uncertainty. On the other hand, if only one intensity $i'$ is likely to be found, then $p_i=\delta_{i,i'}$, and the entropy is zero, representing complete knowledge.

For our computations, we found the entropies shown in Table \ref{tab:entropy}. Background potentials estimated using Pix2Pix showed the smallest average entropy among images, whereas the other two techniques displayed similar entropies. This is probably due to the nature of the estimation processes. Pix2Pix works by, given an SGM image, creating a potential image that minimizes an adversarial loss on a set of theoretically generated images. The model is then used to estimate real images, but there is no guarantee that experimental images have the same probability distribution as that of the theoretical set. This is furthered enforced by a very limited theoretical dataset. CNN is a more direct approach that does not depend explicitly on the probability distributions, rather, it produces a non-linear mapping between SGM and potential images. On the other hand, it is not garanteed that the experimental images have the same mapping as that obtained from the thoeretical set. Finally, evolutionary search works by progressively obtaining a potential that, given a physical model, produces an SGM image that is the close to the experimental one. Consequently, this last technique is not limited by the finite dataset or a specific mapping generated by a finite set. As a result, evolutionary search produces potential backgrounds that are more complex and richer in details.

\begin{table}[h]
	\caption{Image entropies computed for Pix2Pix GAN, Cellular Neural Networks (CNN), and Evolutionary Search (ES)}\label{tab:entropy}
	\begin{tabular}{@{}lccc@{}}
		\toprule
		 & GAN & CNN & ES\\
		\midrule
		Minimum   & 3.11   & 5.60  & 5.94  \\
		Average    & 3.75   & 6.00  & 6.06  \\
		Maximum  & 4.37   & 6.56  & 6.20  \\
		\botrule
	\end{tabular}
\end{table}

\subsection{Potential Roughness}
The root mean square average of the estimated potential, or roughness, is defined as:

\begin{equation}
	R=\sqrt{\frac{1}{L_xL_y}\sum_{\substack{x=1\\y=1}}^{L_xL_y}  \left(P(x,y)-\langle P \rangle\right)^2},
\end{equation} 
where $L_x$ and $L_y$ are the width and height of the image, $P(x,y)$ is the pixel-value at coordinates $(x,y)$, and $\langle P\rangle$ denotes the pixel average of the image.

\begin{figure}[h!]
	\centering
	\includegraphics{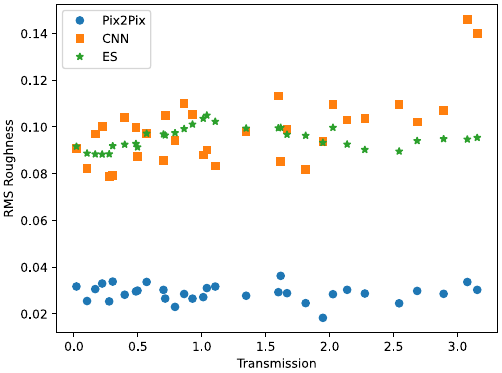}
	\caption{RMS roughness as a function of transmission for potentials estimated via Pix2Pix GAN (blue disks), cellular neural networks (orange filled squares), and evolutionary search (green stars).}
	\label{fig:rough}
\end{figure}

The surface roughness quantifies the short-wavelength spatial variability of the estimated potential, and, consequently, gives an idea of potential texture. Potential roughness plays an important role for quantum coherent devices, since it may cause variations and broadening of the energy levels of confined electrons. Moreover, potential roughness may increase electron scattering and dephasing, which leads to reduced coherence times. These effects can directly affect the scanning gate microscopy patterns. The transmission-dependent roughness measures for the different techniques is shown in Fig. \ref{fig:rough}.

Pix2Pix GAN estimates potentials with little roughness and variability of roughness, which can not correspond to a realistic scenario because as the transmission is reduced, the background potential moves closer to the Fermi level and its effects should be more pronnounced. Potentials estimated by CNN and ES are rougher, and follow a similar trend for transmission values smaller than approximately $2$. The roughnesses of potentials estimated by ES, however, show a smaller variability, which is an expected result. Changing the transmission less than unity should not cause a significant movement of the background potential closer to the Fermi level.

\subsubsection{Fractal Dimension}
The fractal dimension is a related concept to the surface roughness. While surface roughness measures irregularities and deviations from a smooth potential, the fractal dimension measures how a pattern fills space at different scales in a self-similar fashion. The Hausdorff-Besicovitch dimension can be understood through a similar transformation. For example, if amplifying an object by a factor $n$, we obtain $n^D$ copies of the same object, we say that $D$ is its dimension. Mathematically, the Hausdorff-Besicovitch dimension is given by:

\begin{equation}
	D_{HB}=\frac{\log(N)}{\log(n)},
\end{equation}
where $N$ is the number of obtained copies.

For calculating the fractal dimension of images, however, we use the Minkowski-Bouligand dimension. It consists of covering the image with a grid, and, for each box of the grid applying a renormalization measure $\mu$ that indicates whether that box is necessary to cover the pattern under analysis. The most used measure is simply the presence of a filled pixel. Our images, however, are gray-scale. Therefore, we normalized the images as: $M\rightarrow\left(M-\min(M)\right)/\left(\max(M)-\min(M)\right)$ and used a measure $\mu_B=\min\left(1,\sum_{b\in B}b\right)$ that counts the total intensity inside a box. The number of boxes necessary to cover the image at some scale $\epsilon$ gives an idea of the intricacy, or level of detail of the image at that scale. Similarly to the Hausdorff-Besicovitch, the Minkowski-Bouligand dimension is obtained from:

\begin{equation}
	D_{MB}=-\lim_{\epsilon\rightarrow 0}\frac{\log\left(N_\epsilon\right)}{\log(\epsilon)},
\end{equation}
where $N_\epsilon$ is the number of boxes. This dimension is obtained from a linear regression of $\log(N_\epsilon)\times \log(\epsilon)$.

\begin{figure}[h!]
	\centering
	\includegraphics{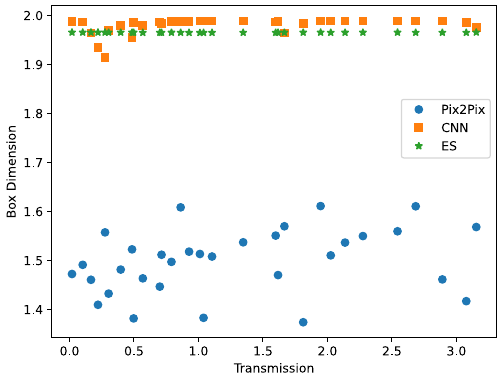}
	\caption{Box dimension as a function of transmission for potentials estimated via Pix2Pix GAN (blue disks), cellular neural networks (orange filled squares), and evolutionary search (green stars).}
	\label{fig:fractal}
\end{figure}

Figure \ref{fig:fractal} shows the Minkowski-Bouligand dimension for the potentials estimated using Pix2Pix GAN, cellular neural networks, and evolutionary seach. While CNN and ES estimate potentials with a fractal dimension close to $2$, Pix2Pix GAN estimates potentials with an average dimension close to $1.5$ with a much greater variability. It is also interesting to note that potentials estimated by ES have a complexity that is nearly constant and does not depend much on the electronic transmission. On the other hand, CNN offers potentials with a similar complexity, but is particularly sensitive to low electronic transmission values.

\subsubsection{Morphological Granulometry}
The estimated potentials display a set of grains that can be studied by a morphological analysis. To perform this computation, first the dilation of an image $M$ is calculated according to a kernel $K$ as:

\begin{equation}
	(M\oplus K)(x,y)=\max_{(i,j)\in K}\left\{M(x+i,y+j)\right\}.
\end{equation}
Similarly, the erosion of the same image is given by:

\begin{equation}
	(M\ominus K)(x,y)=\min_{(i,j)\in K}\left\{M(x+i,y+j)\right\}.
\end{equation}

The opening of the image is then given by the dilation of the erosion of the image:

\begin{equation}
	M\bullet K=(M\ominus K)\oplus K.
\end{equation}
This morphological operation removes structures smaller than the kernel from the image.

The granulometry $G_n(M)$ is then computed by finding the cardinality of the opening of the image at different kernel sizes $n$. The grain size distribution is finally given by the difference of successive granulometries:

\begin{equation}
	S_k(M)=G_k(M)-G_{k+1}(M).
\end{equation}
In all our calculations, we used a square kernel, and used the mean of the elements as the cardinality function. 

\begin{figure}
	\centering
	\includegraphics{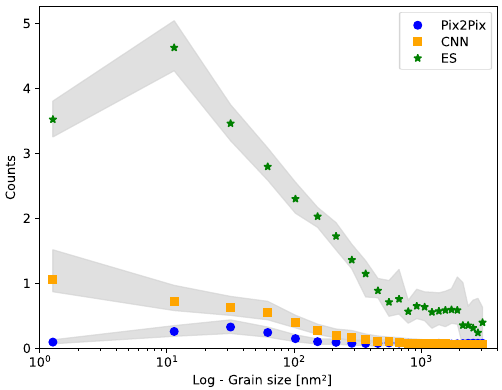}
	\caption{Average grain distribution spectra for the potentials estimated by Pix2Pix GAN (blue discs), cellular neural networks (orange filled squares), and evolutionary search (green stars). The shaded regions cover the minimum and maximum values for each grain size.}
	\label{fig:granu}
\end{figure}

The results obtained with each technique are shown in Fig. \ref{fig:granu}. Our results indicate that the grain distribution exhibits a higher degree of dispersion with a maximum value around $31.64$ nm$^2$, corresponding to a dot with a $3.17$ nm radius. The spectrum for the potential obtained using CNN is also dispersed with a maximum around $1.26$ nm$^2$, corresponding to a $0.635$ nm radius dot. Finally, the spectrum for the potential obtained using evolutionary search peaks around $11.39$ nm$^2$, which corresponds to a $1.90$ nm radius dot. The Bohr radius for InGaAs is around $20$ nm \cite{InGaAs}, which has an area of $1,256.64$ nm$^2$. Although all techniques estimate more granular potentials, the spectrum obtained using evolutionary search demonstrates reduced variation in the vincinity of this value, whereas the spectra obtained with the other two techniques decay to zero.

\subsubsection{Quenched Disorder Potential}
Disorder in the background potential can be influenced by external factors such as the electrostatic field of the side gates. For example, impurities can be charged by the side gate-induced electric field in the channel, or piezoelectric effects can induce strain variations. Also, surface trap states in the dielectric or in the channel can create localized structures that are not quenched. On the other hand, there are some potentials, such as those caused by lattice defects and surface roughness, that are quenched and do not vary due to any external factor. Some charged impurities and the alloy potential caused by the random distribution of different atomic species are also quenched.

By varying the voltages on the side-gates, we can modify the effective aperture of the quantum constriction, leading to changes in the transmission through it. Our challenge arises from the fact that we estimate the potential based on this transmission map, which is influenced by external parameters. This complicates the estimation of the quenched potential features, which ideally should remain unaffected by external parameters. To estimate the quenched potential, we look for features that remain consistent and are minimally affected across different transmissions. 

\begin{figure}[h!]
	\centering
	\includegraphics{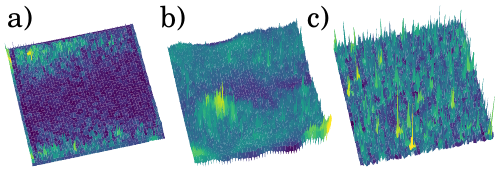}
	\caption{Estimated alloy potential for a) Pix2Pix GAN b) cellular neural networks, and  c) evolutionary search.}
	\label{fig:alloy}
\end{figure}

To achieve this, we compute the pixel-wise precision, which is given by the inverse of the variance, computed for all images along the transmission dimension. This measure is high when there is greater certainty in the data set, and low other otherwise. Figure \ref{fig:alloy} shows the estimted alloy potential obtained for the different techniques.

The three techniques estimate static alloy potentials that are correlated by only $7$ \%. Although they are statistically uncorrelated, both Pix2Pix GAN and CNN try to infer a potential given training examples. This is highlighted by more intense regions on the top and bottom of the images corresponding to the gate regions. This kind of inference assumes that the the SGM image corresponds to a potential, at least, similar to those in the training set. Moreover, the estimated potential may not necessarily reproduce the obtained SGM signal, since there is no physical law conditioning the inference. On the other hand, ES tries to find, through a physical law, the potential that creates an SGM signal closest to the experimentally obtained. However, it requires a physical law, and the inference assumes that the physical phenomenon is strictly modeled by that law. In our case, we are assuming that the SGM signal was produced by a one-electron wavefunction.

\section{Conclusion}\label{sec:conclusion}
 In conclusion, our study examined the estimation of background potentials in two-dimensional electron gases (2DEGs) using three distinct machine learning techniques: Pix2Pix GAN, cellular neural networks (CNN), and evolutionary search (ES). The evaluation encompassed several crucial aspects of the estimated potentials, shedding light on their characteristics and implications.
 
 Entropy analysis revealed that Pix2Pix GAN consistently yielded potentials with the lowest average entropy among the three techniques, while CNN and ES exhibited similar entropy values. This discrepancy can be attributed to the nature of the estimation processes, with Pix2Pix GAN relying on a theoretical dataset that may not perfectly match the experimental data distribution.
 
 Regarding potential roughness and its variation, Pix2Pix GAN produced potentials with limited roughness and variability. This outcome, however, departs from realistic scenarios where the background potential's proximity to the Fermi level should result in more pronounced effects as the transmission decreases. In contrast, CNN and ES generated rougher potentials that align more closely with expected trends, with ES demonstrating a notably reduced variation in roughness.
 
 The examination of fractal dimensions exposed interesting differences. Pix2Pix GAN estimated potentials with an average dimension close to 1.5 and significant variability. In contrast, CNN and ES provided potentials with a fractal dimension near 2. Notably, ES-generated potentials displayed consistent complexity, independent of electronic transmission levels, while CNN exhibited sensitivity to lower transmission values.
 
 Granulometry analysis showcased dispersed grain distributions for all techniques, with peak values indicating granule sizes. The spectra derived from CNN and ES showed dispersion, with peak values corresponding to specific radii. Interestingly, ES demonstrated less variation near the Bohr radius, indicating a narrower distribution.
 
 Finally, an analysis of the quenched potential revealed that Pix2Pix GAN and CNN primarily inferred potentials based on training examples. This inference approach assumes a degree of similarity between the SGM image and those in the training set. However, it does not guarantee the accurate reproduction of the SGM signal, as no physical law governs the inference process. In contrast, ES sought to find a potential that best matched the experimentally obtained SGM signal based on physical laws, relying on a strict modeling assumption.
 
 In summary, our investigation into background potential estimation techniques for 2DEGs using SGM data elucidated the strengths and limitations of each approach. While Pix2Pix GAN and CNN aim at modeling the general relationship between the SGM signal and the potential without access to a conversion function, ES require a physical assumption to guide the search. Therefore ES can operate on a single stance at a time and fine-tune estimate the potential reversely given the SGM signal. 
 Pix2Pix GAN, while producing low-entropy potentials, may not align with realistic scenarios. CNN and ES, by contrast, demonstrated rougher potentials more consistent with expectations. Our findings provide valuable insights into the selection of suitable techniques for addressing inverse problems in this domain, offering guidance for future research and applications in quantum electronics and nanotechnology.

\bibliographystyle{unsrt}
\bibliography{paper}

\end{document}